\documentclass[aps,prb,superscriptaddress,
  showpacs,floatfix
 ,twocolumn
]{revtex4-1}

\usepackage{times}
\usepackage{amsmath,bm,amsfonts}
\usepackage{graphicx}
\usepackage{color}

\begin{document}

\title{Inevitable high density of oxygen vacancies on the surface of 
LaAlO$_3$/SrTiO$_3$ heterostructures}

\author{Yun Li}
\email{Email: yun.li141@gmail.com}
\affiliation{Department of Physics and Electonic Engineering Hanshan Normal
Univerisity, Chaozhou 521041, China}
\affiliation{Department of Physics and Astronomy, 
Seoul National University, 
Seoul 151-747, Korea}
\author{Xinyuan Wei}
\affiliation{State Key Laboratory of Surface Physics, Department of Physics, 
Fudan University, Shanghai 200433, China}
\author{Jaejun Yu}
\affiliation{Department of Physics and Astronomy, 
Seoul National University, Seoul 151-747, Korea}


\begin{abstract}
Using density-functional-theory (DFT) calculations with the HSE06 hybrid 
functional, we accurately evaluate the critical thickness of LaAlO$_3$ film for 
the intrinsic doping in LaAlO$_3$/SrTiO$_3$ 
(LAO/STO) heterstructures. The calculated critical thickness of 6 
unit-cell (uc) layers  suggests to rule
out the intrinsic doping mechanism. 
We also calculate the density of oxygen vacancies on the 
LAO surface at varying LAO thicknesses, preparation oxygen pressures and 
temperatures by using the condition of chemical equilibrium and DFT 
calculations. We find that once  LAO 
thickness $\geq$3 uc high-density ($\sim10^{14} cm^{-2}$) oxygen 
vacancies will inevitably exist on the LAO surface of the LAO/STO 
heterstructures     
even though the samples are grown under high oxygen pressure. The 
oxygen vacancies are stabilized by releasing the electrostatic energy in the 
LAO film.
\end{abstract}


\maketitle

Perovskite transition metal oxides have rich functionalities and have been a 
hotspot in the area of material science. In recent years, heterostructures of   
complex perovskite oxides made this area more colorful and appealing. Some 
intriguing properties, which are absent in the individual components, may 
emerge in the heterostructures. The polar-nonpolar heterostructure 
LAO/STO is a  typical example. 
LAO and STO are both wide-gap band insulators, 
but a high-mobility two-dimensional electron gas (2DEG) was found to exist 
at the  LaO/TiO$_2$ interface. \cite{Ohtomo2004}
Moreover, ferromagnetism, 
\cite{Brinkman2007,Kalisky2012,Bi2013Room}
supercondctivity,\cite{Reyren2007b,Caviglia2008c,Li2011c,Bert2011} 
metal-to-insulator transition 
(MIT),\cite{Thiel2006,Cen2008b} were successively found 
at the interface.
These  properties closely relay on the interfacial carriers. The polarity of 
the LAO film plays a key role in the formation of the carreirs. 

\begin{figure}[htbp] 
\centering
\includegraphics[width=0.4\textwidth]{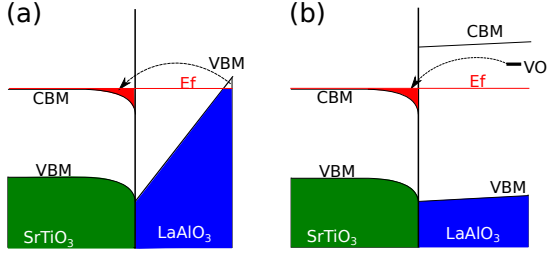}
\caption{\label{Fig1} Band diagrams of (a) intrinsic doping and (b) 
surface oxygen-vacancy doping in the LAO/STO heterostructures.}
\end{figure}

In defect-free LAO/STO heterostructures,   a polar 
electric field will appear in the LAO film along [001] direction. 
With increasing LAO thickness the electrostatic energy accumulated in the LAO 
film will increase accordingly, finally leading to the electronic 
reconstruction as proposed by Nakagawa et al, i.e., charge transfers from the 
LAO surface into 
the interface at the STO side.\cite{Nakagawa2006}
Based on this picture two mainstream doping mechanisms, i.e. 
intrinsic doping and oxygen-vacancy doping,  were proposed.
As shown in Figs.~\ref{Fig1} (a) and (b), both doping 
mechanisms are triggered by the polar field in the LAO film. 
Usually, it was believed that the intrinsic doping 
occur in the LAO/STO heterostrucutes prepared under 
high oxygen pressures because high oxygen pressure could easily remove oxygen 
vacancies in oxides.  
Several DFT calculations showed that in defect-free 
LAO/STO there is a striking band gradient in the LAO film and a critical LAO
thickness of 4 uc layers for the intrinsic doping, at which the valence band 
maximum  (VBM) of the LAO surface exceeds the conduction band minimum (CBM) of 
the STO.\cite{lee2008charge,pentcheva2009avoiding,son2009density,Li2010j} 
Meanwhile, several experiments also 
reported that the LAO/STO heterostrucutes  prepared under 
high oxygen pressures present a critical thickness of 4 uc for conducting 
interface, seemingly conforming the intrinsic 
doping.\cite{Thiel2006,Bell2009,Segal2009,Takizawa2011,Kalisky2012}
 However, the expected large band gradient, as the 
necessary and direct evidence of intrinsic doping, has not yet been observed, 
in contrast, the observed values are 
negligible.\cite{Segal2009,Takizawa2011,Slooten2013,Drera2013}
Moreover, increasing experimental studies have shown that for the 
LAO/STO heterostructures prepared under high oxygen pressures even at the LAO  
thickness of 3 uc layers  the localized carriers already appear at the 
interface.\cite{Savoia2009c,Takizawa2011,Drera2011a,Pfaff2018}  Noticeably, the 
above DFT studies, which performed the calculations 
with Local-density-approximation (LDA) or generalized-gradient-approximation 
(GGA) functional, largely underestimated the band gap of STO 
($\sim$ 1.8 eV in the calculations, while 3.2 eV in experiment). This implies 
that the calculations underestimated the critical thickness of the intrinsic 
doping. Thus, the intrinsic doping is doubtable.
On the other hand,  theoretical studies on oxygen vacancies showed that oxygen 
vacancies may easily exist on the surface of LAO/STO heterostructures because 
the charge transfer from the LAO surface to the interface could release the 
electrostatic energy accumulated in the LAO film.\cite{Zhong2010d,Li2011f} 


Nowadays, the polar-nopolar heterostructures like LAO/STO have been 
extensively used to tune the electronic properties of perovskite oxides. 
However, the doping mechanism in LAO/STO heterostructures is still uncertain. 
In this paper, we 
study the intrinsic and oxygen-vacancy doping mechanisms in LAO/STO 
heterostructures. First, we accurately calculate the critical thickness of LAO 
film for the 
intrinsic doping by using density-functional-theory (DFT) calculations with the 
HSE06 hybrid functional. The calculated value of 6 uc implies that 
 the intrinsic doping mechanism is ruled out. Then, we derive the density of 
oxygen vacancies on the 
LAO surface at varying LAO thickness, preparation oxygen pressures and 
temperatures by using the condition of chemical equilibrium and DFT 
calculations.  We find that once  LAO 
thickness $\geq$3 uc high-density ($\sim10^{14} cm^{-2}$) oxygen 
vacancies will inevitably exist on the LAO surface of the LAO/STO 
heterstructures.   


To correctly evaluate the critical LAO thickness for intrinsic doping in  the 
defect-free LAO/STO heterostructures, we carried out DFT calculations with the 
HSE06 hybrid functional  by using VASP 
code.\cite{kresse1996efficient,kresse1999ultrasoft,blochl1994projector}  
The HSE06 hybrid functional,\cite{Heyd2003Hybrid,Heyd2004Efficient}  
which contains 23\% of the exact exchange and 77\% of 
the PBE exchange energy,\cite{PhysRevLett.77.3865} correctly 
estimates  the band gap of bulk STO (3.2 eV). The 
heterostructures were modeled in c(2$\times$2) supecells, 
which consist of 4 uc layers of STO, 
n (n=4, 5, 6) uc layers of LAO and about 14 {\AA} of 
vacuum region. The atoms in the bottom two STO layers were fixed at their 
bulk positions and the others were fully relaxed until the forces on atoms were 
less than 0.02 eV/{\AA}. The dipole correction
method was used to correct the spurious electrostatic
field induced by the periodic boundary condition and the potential difference 
between two different terminations of the 
slabs.\cite{makov1995periodic} 

\begin{figure}[htbp] 
\centering
\includegraphics[width=0.4\textwidth]{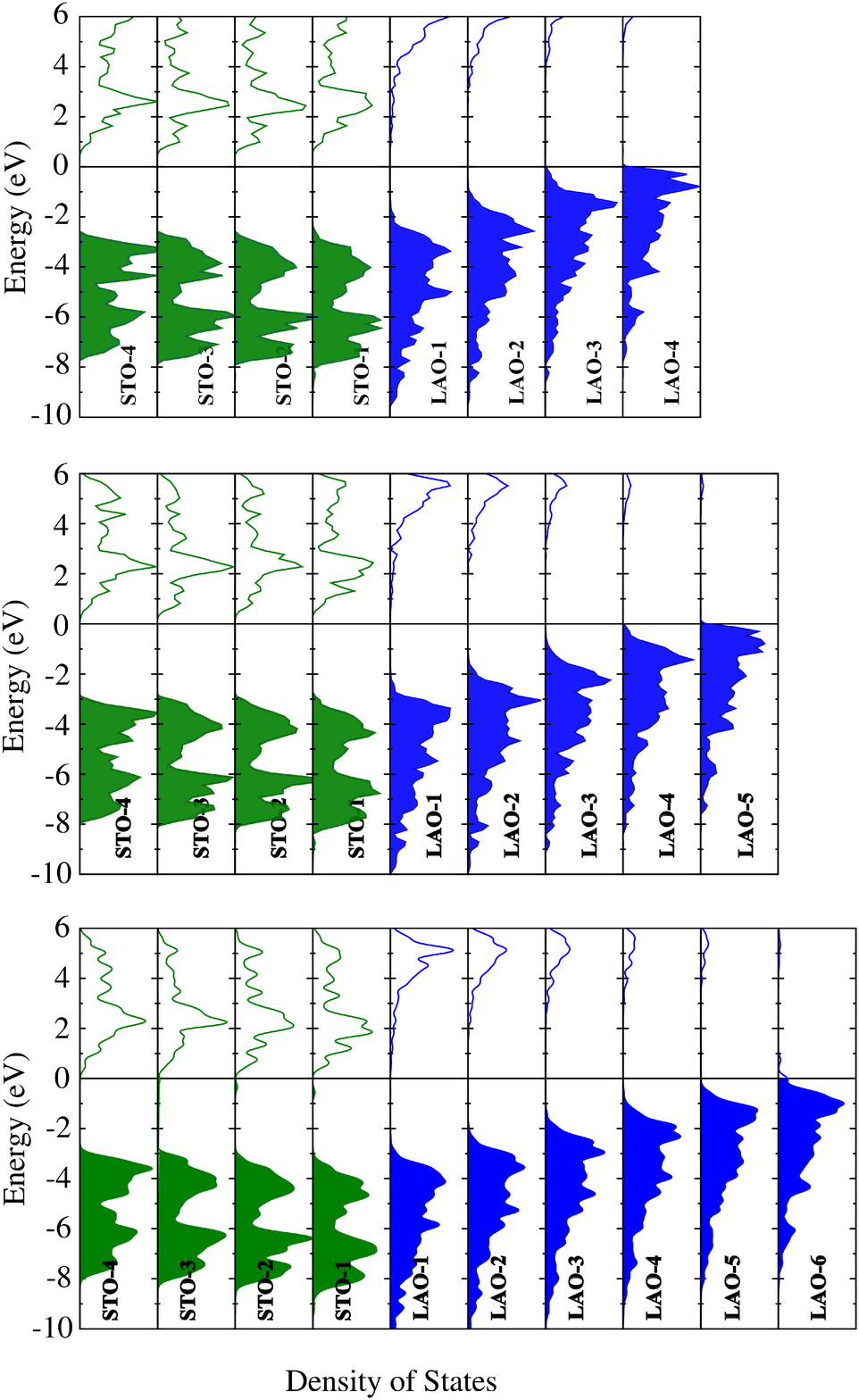}
\caption{\label{Fig2} Layer-resolved densities of states  for defect-free 
LAO/STO heterostructures with n (n=4, 
5, 6) uc layers of LAO. }
\end{figure}

Figure~\ref{Fig2} shows the layer-resolved densities of states (LDOSs) of 
the LAO/STO heterostructures with n (n=4, 5, 6) uc layers of LAO. For the 
4-uc heterostructure there is still a gap of about 0.6 eV between the LAO VBM 
and the STO CBM. The gap is finally closed at 6 uc. This result clearly 
indicates that for defect-free LAO/STO heterostructures the LAO 
critical thickness for the intrinsic doping is 6 uc, 
instead of 4 uc observed in experiments. And considering  
the observed negligible band-shifts in the 
LAO layers, we suggest that the 
intrinsic doping mechanism is basically ruled out.  


Usually, oxygen vacancies hardly exist in bulk LAO under a high oxygen pressure 
condition because the binding energy of an oxygen atom in LAO bulk 
($\sim$ 11 eV) is far higher than  in oxygen gas ($\sim$ 5 eV).
In contrast to  bulk LAO, oxygen vacancies could easily form on the LAO surface 
of the LAO/STO heterostructures because the transfer of  oxygen-vacancies 
induced charge from the LAO 
surface into the interface could partially  compensate the polar 
field, therefore lowering the electrostatic energy as shown in Fig.~\ref{Fig1} 
(b). The energy gain arising from lowering the electrostatic 
energy compensates the binding-energy cost of the surface oxygen 
vacancies. Previous DFT studies have shown that the formation energy of 
an oxygen vacancy on the surface of the LAO/STO heterostructures could be a few 
eV lower than in bulk LAO, and oxygen vacancies 
favor to appear 
on the surface of the LAO/STO heterostructures.\cite{Zhong2010d,Li2011f}

\begin{figure}[htbp] 
\centering
\includegraphics[width=0.45\textwidth]{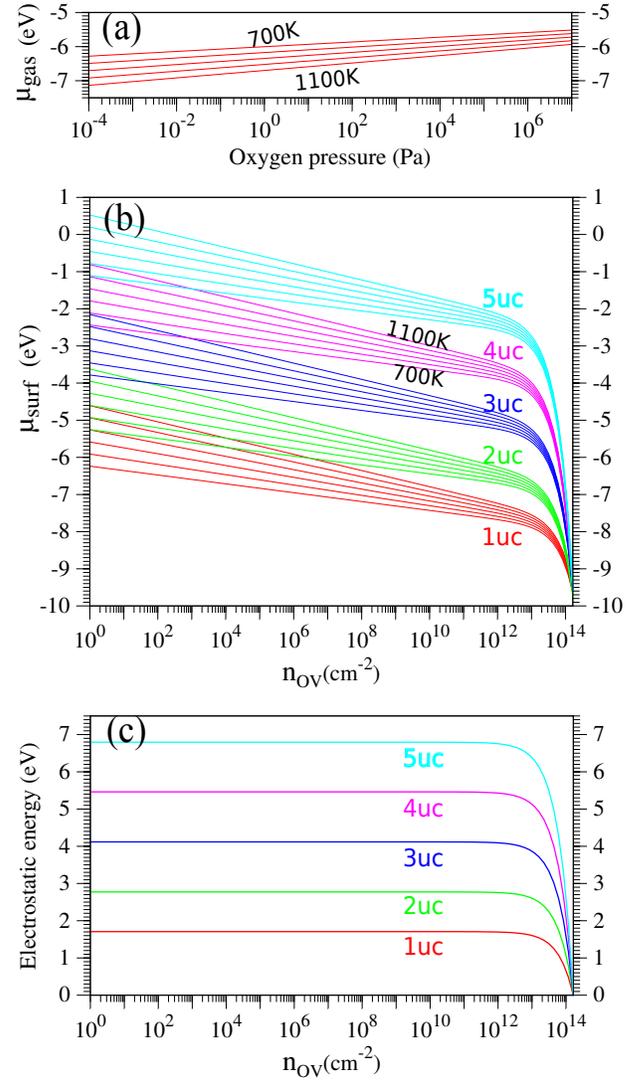}
\caption{\label{Fig3}(a) Chemical potential of an oxygen atom in
O$_2$ gas as a function of pressure and temperature. (b) Chemical 
potential of an oxygen atom at the surface  as a 
function of the density of oxygen vacancies, LAO thickness and 
temperature. The temperature varies in 100-K steps between 700K and 1100K for 
the curves of every thickness.  (c) The increment of electrostatic energy by 
adding an oxygen atom to fill up an oxygen vacancy at the LAO surface of the 
LAO/STO heterostructures as a function of the density of oxygen vacancies and 
LAO thickness. }
\end{figure}

A quantitative relation between the 
density of surface oxygen vacancies and the LAO thickness, oxygen pressure, 
temperature can be established by using the chemical equilibrium condition. For 
 grown  LAO/STO heterostructues  in chamber with the oxygen pressure $P$ and 
temperature $T$, under chemical equilibrium  the chemical potential of the 
surface  oxygen atoms on the surface of the heterostructues 
$\mu_{surf}$ is equal to the chemical 
potential of oxygen 
atoms in the oxygen gas $\mu_{gas}$. 
Using the method in 
Ref.[\onlinecite{Guhl2010}] $\mu_{gas}$ can be expressed as
\[ \mu_{gas}(T,p)=\frac{1}{2}[E_{\rm O_2}+\mu_{\rm O_2}(T,p^{\circ})+k_{\rm
B}Tln(\frac{p}{p^{\circ}})], \]
where $E_{\rm O_2}$ is the binding energy of a gaseous oxygen molecule,  
$\mu_{\rm O_2}(T,p^{\circ})$ is relative chemical potential of gaseous oxygen 
molecules at the standard pressure $p^{\circ}$=1 bar and temperature $T$.  
$E_{\rm O_2}$ was obtained from the first-principles calculation, and $\mu_{\rm
O_2}(T,p^{\circ})$ was obtained by following the method  in 
Ref.[\onlinecite{Guhl2010}]. 
Figure~\ref{Fig3}(a) displays $\mu_{gas}$, in which the ranges of oxygen 
pressure and temperature used in previous 
experiments are covered. $\mu_{gas}$ can vary from 
-7.1 eV (at low pressure and high temperature) to -5.4 eV (at high pressure and 
low temperature).  



For a LAO/STO heterostructure with $N_{OV}$ oxygen vacancies on the surface, 
$\mu_{surf}$ can be written in the form (see Supplementary information) 
\[
\mu_{surf}=\frac{\partial(E_0-E_{OV})}{\partial N_
{OV}}+3k_BT+T\frac{\partial S}{\partial N_{OV}} .
\]
The chemical potential is contributed by three parts, the changes of total 
energy, thermal vibration energy, and entropy's term, which are denoted as 
$\mu_{surf}^{te}$, $\mu_{surf}^{tv}$, and $\mu_{surf}^{en}$, respectively.
$\mu_{surf}^{en}$ can be approximately derived as 
$\mu_{surf}^{en}=k_BTln(n_O-n_{OV})/(n_{OV})$,
where $n_O$ is area density of oxygen atoms on the surface AlO$_2$ layer 
and $n_{OV}$ is area density of the oxygen vacancies.
To derive $\mu_{surf}^{te}$ we resort to a capacitor model combined with DFT 
calculations. The total energy is decomposed  into two parts,  the
binding energy of oxygen atoms on the surface of LAO film without the polar 
field and the electrostatic energy induced by the polar field in the LAO film. 
Considering that the LAO film with the surface and interface is a 
parallel-plate capacitor with thickness 
$d$ and area  $A$, and  the oxygen 
vacancies discharge the capacitor, so the total 
energy difference can be expressed as 
\[
E_0-E_{OV}
=\frac{d}{2\epsilon_0\epsilon_r A}[Q_0^2-(Q_0-2eN_{OV})^2]-N_{OV}E_b
\]
where $\epsilon_r$ is the relative dielectric constant of the LAO 
film, $E_b$ is the 
binding energy of an oxygen atom at the LAO surface without the polar field, 
$Q_0$ is the total polar charge on the surface without oxygen vacancy, the 
factor of  $-2e$ before $N_{OV}$ means 
that one oxygen vacancy discharges the capacitor by a charge of two electrons.
Thus, 
$$
\mu_{surf}^{te}
=-E_b-\frac{de^2}{2\epsilon_0\epsilon_r}(8n_{OV}-4n_0),
$$
where $n_0$ ($n_0e=Q_0/A$) is the 
density of polar charge at the 
surface, i.e. 0.5$e$ per uc area. 
First-principles calculations with supercell method can give the total 
energies of  supercells without oxygen vacancy and with  various densities of 
oxygen vacancies. 
By fitting the formula 
$E_0-E_{OV}$ we could extract the two parameters, $E_b$ and $\epsilon_r$, and 
then obtain the chemical potential $\mu^{surf}_{te}$ as function of  the 
LAO thickness and density of oxygen vacancies. The total-energy calculations 
were carried out with the VASP 
code (see detailed method and calculated data in Supplementary information). 
Four densities of oxygen vacancies, one vacancy in (2$\times$2), 
(3$\times$2), (3$\times$3), and (4$\times$3) supecells respectively,  were 
calculated.

Figure~\ref{Fig3}(b) displays $\mu_{surf}$, which increases remarkably with 
increasing LAO thickness  except in the vicinity of the high end of the density 
of oxygen vacancies. Figure~\ref{Fig3}(c) shows the  
increment of electrostatic energy by adding one oxygen atom to fill up a 
surface oxygen vacancy, i.e., the electrostatic energy released by an oxygen 
vacancy. In the region of low density of oxygen vacancies the energy increases 
largely with increasing LAO thickness  because the thick LAO 
film accumulates more electrostatic energy than the thin film. While in the 
vicinity of the high end of the density of oxygen vacancies the energies for 
different thicknesses  converge to zero because the polar filed has been 
compensated and more oxygen vacancies do not release electrostatic energy any 
more.      
Comparing with $\mu_{gas}$ (-7 $-$ -5 eV), $\mu_{surf}$
in a large region of the density of oxygen vacancies 
is higher, indicating that surface oxygen vacancies will definitely exist in 
the hererostrucutes with thicker LAO film. 

\begin{figure}[htbp] 
\centering
\includegraphics[width=0.45\textwidth]{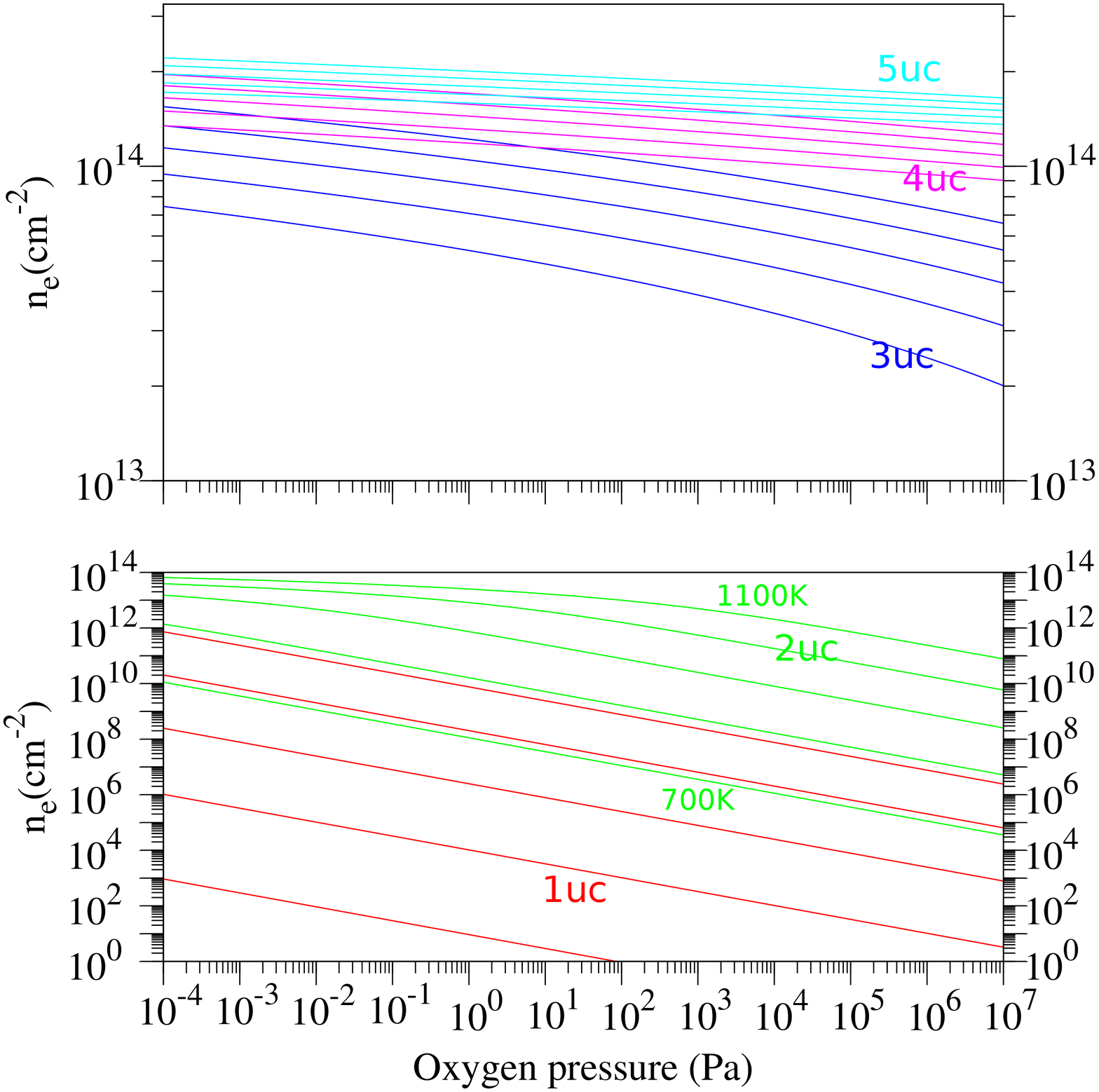}
\caption{\label{Fig4} Density of oxygen-vacancy-induced carriers at 
the LAO/STO interface as a function of LAO thickness,  oxygen 
pressure  and temperature. The temperature varies in 100-K steps between 700K 
and 1100K for the curves of every thickness. Lower panel for d=1, 2 uc, upper 
panel for d=3, 4, 5 uc. }
\end{figure}

Using the relation $\mu_{gas}=\mu_{surf}$ under the condition of 
chemical equilibrium we derive  the density of oxygen-vacancy-induced carriers 
at the LAO/STO interface  and plot it in Fig.~\ref{Fig4}. For all LAO thickness 
the carrier densities decrease with increasing the oxygen 
pressure or decreasing the temperature.  For $d=1,2$ uc the carrier densities 
change sensitively with 
varying the oxygen pressure and temperature, and the carrier densities could be 
negligible at high pressure and low temperature. In contrast,  for $d\geq3$ 
uc the carrier densities change slowly with the oxygen pressure and 
temperature, and keep high values ($2\times10^{13}-2.2\times10^{14} cm^{-2}$) 
even under a very high pressure ($>$1 bar).    
This result evidently 
shows that for the LAO/STO with over 3 uc layers of LAO a high density of 
oxygen vacancies  about  $10^{14} cm^{-2}$ inevitably exist at the LAO surface 
even through a high oxygen pressure is applied to sample preparation. 
Noticeably, with increasing LAO thickness the density of  
carriers will saturate at 6 uc  because according to the intrinsic doping 
picture the electrostatic energy in the defect-free LAO film will reach to 
the maximum. And the maximum carrier density in the LAO/STO  
heterostructures grown under high oxygen pressures could not research to the 
value of 0.5 electron per uc area, i.e. $3.3\times10^{14} cm^{-2}$. As shown 
in Fig.~\ref{Fig3}(c), the surface oxygen vacancies on the LAO surface are 
mainly stabilized by releasing the electrostatic energy. Whereas at this carrier 
density $\mu_{surf}$ ($\sim$ -10 eV) is far lower than $\mu_{surf}$, implying 
that oxygen atoms will flow into LAO to fill up the vacancies.   

%


Noticeably, under high oxygen pressure and low temperature the carrier 
densities for $d=1, 2$  uc and $d\geq3$ uc present a remarkable difference. In 
the  former the carrier density could be negligible or very low, while in the 
later high-density carriers always exist.
Moreover, the cation intermixing, which was observed to exist in 2 uc layers at 
the interface,\cite{Choi2012} could further reduce the polar field in the LAO 
film and might lower the the carry density of $d=1, 2$ uc  to an undetectable 
level. 
This is in very good agreement 
with recent experimental results obtained by using resonant inelastic soft 
x-ray scattering, in which no interfacial carrier was observed until 3 uc layer 
of LAO.\cite{Pfaff2018} Several previous experimental studies also presented 
the same evidence.\cite{Savoia2009c,Takizawa2011,Drera2011a,Pfaff2018} It should 
be noted that the carriers observed in the LAO/STO with 3 uc 
layers of LAO are localized and has no contribution to 
conduction.
In addition, since the high-density 
carriers induced by the surface oxygen vacancies  compensate the main part of 
the polar field, the band gradient in the LAO film should be quit small. This 
is consistent with the small band gradient observed in the experiments.




In conclusion, 
our hybrid-functional DFT calculations clearly show that in defect-free LAO/STO 
heterostructures the intrinsic doping occurs at the LAO thickness of 6 
uc layers, instead of 4 uc layers. This indicates that the observed 
metal-insulator transition at 4 uc LAO layers is not induced by the intrinsic 
doping. On the contrary, our theoretical study on the density of  
oxygen vacancies show that the intrinsic doping never happened, instead oxygen 
vacancies on the LAO surface produce the carriers because the 
oxygen vacancies are stabilized by releasing the electrostatic 
energy in the LAO film. Once LAO thickness $\geq$ 
3 uc layers, a high density of oxygen vacancies inevitably appears on the 
surface even through the heterostructure is prepared under a very high oxygen 
pressure. Accordingly, a high density of carriers is generated at the 
interface, and it varies dominantly with the LAO thickness, and slowly with 
oxygen pressure and temperature.     
  

\acknowledgments 
This work was supported by the Scientific Research
Foundation for the Returned Overseas Chinese Scholars of
State Education Ministry (Grant No. [2015]-1098) and  the Open Project 
of State Key Laboratory of Surface Physics of Fudan University.


%

\newpage
\section{Supplementary information}

For a LAO/STO heterostructure with $N_{OV}$ oxygen vacancies on the surface, 
the Gibbs free energy $G_{OV}$ is 
\begin{equation*}
\begin{aligned}
 G_{OV}&=U_{OV}-TS=G_0+(U_{OV}-U_0)-TS \\
 &=G_0+(E_{OV}-E_0)-3k_BTN_{OV}-TS ,
\end{aligned}
\end{equation*}
where $U_{OV}$ and $U_0$ are the internal energies of the LAO/STO 
heterostructure 
with oxygen vacancies and without oxygen vacancies, respectively, S is the 
configuration entropy of oxygen vacancies and it is zero 
for the LAO/STO heterostructure without oxygen vancancy, $G_0$ is the Gibbs 
free energy of the
heterostructures without oxygen vacancy and it is equal to $U_0$. The 
internal energy $U_{OV}$ or $U_0$ comprises of two parts, the total energy at 
zero temperature $E_{OV}$ or $E_0$ and the thermal vibration energy. The item 
$-3k_BTN_{OV}$ is  the difference of thermal vibration energy between the 
heterostructure without oxygen vacancy and with $N_{OV}$ oxygen vacancies. 
The $\mu_{surf}$ follows from 
\begin{equation*}
\begin{aligned}
\mu_{surf}&=-\frac{\partial G_{OV}}{\partial N_{OV}}\\ 
&=\frac{\partial(E_0-E_{OV})}{\partial N_
{OV}}+3k_BT+T\frac{\partial S}{\partial N_{OV}} .
\end{aligned}
\end{equation*}
The configuration entropy of the oxygen vacancies
can be evaluated approximately as
$$ S=k_BlnW=k_Bln\frac{N_O!}{(N_O-N_{OV})!N_{OV}!} ,$$
where $N_O$ is the total number of oxygen sites at vacancy-free surface AlO$_2$
layer.  
$$ \mu_{surf}^{en}=T\frac{\partial S}{\partial 
N_{OV}}=k_BTln\frac{N_O-N_{OV}}{N_{OV}}=k_BTln\frac{n_O-n_{OV}}{n_{OV}} .$$


The total-energy calculations were carried out with the VASP 
code.
The generalized gradient approximation 
for the exchange-correlation functional was applied to 
evaluated the electron-electron interaction.\cite{PhysRevLett.77.3865} The 
calculations were performed 
with the projector augmented wave method,\cite{blochl1994projector} in which 
the cutoff kinetic energy for the plane wave basis set is 450 eV.
The LAO/STO heterostructures were modeled in (2$\times$2), 
(3$\times$2), (3$\times$3), and (4$\times$3) supecells.
The total energies of the supercells without oxygen vacancy and with an oxygen 
vacancy on the LAO surface were calculated. 
The supercells consist of 4 uc layers of STO, 
n (n=1,2,3) uc layers of LAO and about 14 {\AA} of 
vacuum region. The atoms in the bottom two STO layers were fixed at their 
bulk positions and the others were fully relaxed until the forces on the atoms 
were less than 0.02 eV/{\AA}. The dipole correction
 was used to correct the spurious electrostatic
field induced by the periodic boundary condition and the potential difference 
between two terminations of the slabs.\cite{makov1995periodic}

\begin{table}
\caption{\label{table1} $E_{OV}-E_0$ (eV) for the LAO/STO heterostructures with 
various LAO thicknesses $d$ (uc), in which $E_0$ and  $E_{OV}$ are the total 
energies of the LAO/STO supercell  without oxygen vacancy and  with one oxygen 
vacancy at the LAO surface, respectively. $E_b$ is the 
binding energy of an oxygen atom on the LAO surface without the polar field.}
 \begin{tabular}{p{1.5cm}p{1cm}p{1cm}p{1cm}p{1cm}p{1.5cm}}
 \hline\hline
Supercell & (2$\times$2) & (3$\times$2) & (3$\times$3) & (4$\times$3) & $E_b$ 
(eV)  \\ 
\hline
  $d=1$   & 8.975 & 8.874 & 8.660 & 8.347 & 9.900\\ 
  $d=2$   & 8.525 & 8.227 & 7.952 & 7.515 & 9.981\\
  $d=3$   & 7.699 & 7.263 & 6.788 & 6.237 & 9.856\\ 
   \hline\hline
  \end{tabular}
\end{table}

Table~\ref{table1} lists the total energy differences $E_{OV}-E_0$. By fitting 
with the formula of $E_0-E_{OV}$ we derived the $E_b$ and $\epsilon_r$, thus 
we obtained $\mu^{surf}_{te}$ for $d=1,2,3$ uc. As mentioned above, in the 
calculations with HSE06 functional the intrinsic doping does not 
occur in the LAO/STO until 6 uc layers of LAO, therefore our
 model is also validate for the LAO/STO with the 4 and 5 uc LAO. We decomposed 
the 3 uc layers of LAO into an interface uc layer, a
middle uc layer and a surface uc layer, and derived their dielectric 
constants respectively. We consider that the LAO films of 4 and 5 uc layers 
just have one 
and two more middle uc layers compared with the film of 3 uc layers, and  
the binding energy  $E_b$ for the 4 and 5 uc is the same as that for 3 uc. 
By this means we derived the  $\mu^{surf}_{te}$ for the LAO/STO 
heterostructures with $d=1, 2, 3, 4, 5$ uc layers of LAO.    
\end{document}